\documentclass{jfm}
\usepackage{xcolor}
\usepackage{graphicx}
\usepackage{natbib}
\usepackage{amsmath}
\usepackage{url}

\newcommand{\edit}[1]{{#1}}

\shorttitle{Water entry of a simple harmonic oscillator}
\shortauthor{J. T. Antolik, J. L. Belden, N. B. Speirs, D. M. Harris}
\title{\edit{Slamming forces during} water entry of a simple harmonic oscillator}

\author{John T. Antolik\aff{1},    
        Jesse L. Belden\aff{2},
        Nathan B. Speirs\aff{2,3},
        \and Daniel M. Harris\aff{1}\corresp{\email{daniel\_harris3@brown.edu}}}

\affiliation{
    \aff{1} Brown University, Center for Fluid Mechanics and School of Engineering, 184 Hope St., Providence, RI 02912, USA
    \aff{2} Naval Undersea Warfare Center Division Newport, 1176 Howell St., Newport, RI 02841, USA
    \aff{3} Brigham Young University, Department of Mechanical Engineering, 350 Engineering Building, Provo, UT 84602, USA}
 
\begin{document}

\maketitle
   
\begin{abstract}
When a blunt body impacts an air-water interface, large hydrodynamic forces often arise, a phenomenon many of us have unfortunately experienced in a failed dive or ``belly flop.’’  Beyond assessing risk to biological divers, an understanding and methods for remediation of such slamming forces are critical to the design of numerous engineered naval and aerospace structures.  Herein we systematically investigate the role of impactor elasticity on the resultant structural loads in perhaps the simplest possible scenario: the water entry of a simple harmonic oscillator.  Contrary to conventional intuition, we find that ``softening’’ the impactor does not always reduce the peak impact force, but may also increase the force as compared to a fully rigid counterpart.  Through our combined experimental and theoretical investigation, we demonstrate that the transition from force reduction to force amplification is delineated by a critical ``hydroelastic’’ factor that relates the hydrodynamic and elastic timescales of the problem.
\end{abstract}

\section{Introduction}

Water entry of solid bodies has been a subject of intense investigation for over a century, with rich multi-scale physics revealed at all stages of the process.  Progress in the field has been principally driven by a need for understanding the hydrodynamic loading experienced by impacting engineered naval structures such as ships, seaplanes, or projectiles, directly motivating early theoretical developments in the area by \cite{von1929impact} and \cite{wagner1932stoss}.  Other impactors such as aerospace structures \citep{seddon2006review} or amphibious autonomous vehicles \citep{siddall2014launching,shi2019numerical} have benefited from advancements in the area.  Beyond informing engineering applications, such forces can prove fatal for human divers if the hydrodynamics are not respected \citep{pandey2022slamming}.  

For relatively blunt bodies such as shallow wedges or spheres, the highest impact forces occur during the very early times of impact, in the so-called “slamming” phase \citep{shiffman1945sphere2,may1970review,moghisi1981experimental,korobkin1988initial,miloh1991initial,howison1991incompressible,abrate2011hull}.  It is well established that the primary contribution to this initial hydrodynamic resistance stems from the added mass effect of the fluid: an appreciable volume of fluid must be accelerated in a short time frame to match the speed of the impinging body \citep{abrate2011hull,truscott2014water,jung2021swimming}.  High impact forces can result in structural damage, present risk to sensitive onboard equipment, or be dangerous for passengers or biological divers.  Thus understanding the relationship between the impactor properties and the resultant impact forces, and developing predictable and controllable ways to mediate such forces, are of utmost importance.

With the motivation of force reduction in mind, many prior works have focused on how the impact forces are influenced by impactor geometry.  For instance, a finely tapered impactor significantly reduces the initial impact forces as compared to more blunt geometries \citep{baldwin1971vertical,bodily2014water,vincent2018dynamics}, a physical principle that biological divers such as seabirds evidently exploit to survive high-speed impacts in pursuit of prey \citep{chang2016seabirds,sharker2019water}.  In many cases, the impactor geometry cannot be suitably modified and other countermeasures must be considered.  For instance, by preceding a primary impactor with a fluid jet or small solid object, some of the underlying liquid can be accelerated or displaced before impact and impact forces of the trailing body notably reduced \citep{speirs2019water,rabbi2021impact}. 
 In terms of modifications to the impactor, a sacrificial and permanently deformable nose cap can be added to absorb some of the energy during impact \citep{shi2019design,li2021dynamic}, but is only effective for a single impact. An alternative approach, common to countless other examples, is to introduce elastic compliance to ``cushion’’ the impact, thereby extending the timescale of the impulse and reducing peak forces.  In the current context, this manifests as a fluid-structure interaction (FSI) problem wherein the structural and hydrodynamic responses are intrinsically coupled.  The role of impactor elasticity in air-water entry has received some limited attention, primarily over the past few decades.

The impact of elastic structures on fluid interfaces are often referred to as “hydroelastic” problems.  A key non-dimensional parameter that naturally emerges in many such investigations is a ratio of timescales sometimes referred to as a hydroelastic factor ($\textrm{R}_\textrm{F}$): the timescale of the hydrodynamic loading to the free fundamental oscillation period of the elastic structure \citep{kim1996coupled,faltinsen1999water,ren2021kinematic}.  The vast majority of prior works have focused on the coupled response of continuously deformable flexible wedges as a model for ship hulls \citep{faltinsen1999water,abrate2011hull,maki2011hydroelastic,panciroli2012hydroelasticity,khabakhpasheva2013elastic,shams2017hydroelastic,ren2021kinematic}, although a few other continuous structures have been studied as well such as elastic spheres \citep{hurd2017water,yang2021hydroelastic}.  Other investigations have proposed simpler lumped mass models (reduced degrees of freedom) of continuous elastic structures in an attempt to simplify the problem and better interpret the consequences of elasticity on the resultant structural forces \citep{miller1951comparison,gollwitzer1995repeated,kim1996coupled,lafrati2000hydroelastic,carcaterra2004hydrodynamic,bogaert2007design}.  Despite these efforts, very few controlled experiments on simplified structures have supported such studies, and none present a systematic exploration of the parameter space.  

 The most similar experimental work to the present study was completed very recently wherein the loading on an axisymmetric two degree-of-freedom (DOF) (one elastic mode) impactor was considered \citep{wu2020water}.  The impactor was composed of a rigid hemispherical nose and slender body connected by a coil spring element.  Only one geometry and spring constant were explored in the work, and for that case, the impact force on the body was reduced compared to a rigid counterpart over all impact velocities tested.  While the finding conforms to standard intuition one might associate with a “cushioned” impact, no predictive model was developed to quantify or generalize the measured effect. As we demonstrate in the present work through combined experiment and modeling, the impact force on the trailing body of a 2DOF elastic system is highly sensitive to both the elastic and hydrodynamic parameters of the problem, and the force can either decrease or increase as a consequence of the elasticity, in general. 

 In the present work, we design and test a 2DOF (corresponding to one axial elastic mode) slender axisymmetric impactor with a hemispherical nose.  By using a configuration of custom flexures as the compliant elements interfacing the nose and body, a highly linear elastic response is achieved without static or sliding friction and only very weak material damping.  Furthermore, the geometry of the overall structure is carefully designed to separate the frequency of the fundamental (axial) mode from all other elastic modes, ultimately rendering it an excellent approximation of a linear 2DOF system.  The deceleration of the body during water entry is directly measured using an onboard untethered accelerometer at very high sampling rate.  A range of impact velocities, spring stiffnesses, nose radii, and nose to body mass ratios are tested and the peak forces measured in all experiments are collapsed along a single curve using inertial scaling and an appropriately defined hydroelastic factor. A critical hydroelastic factor defines the transition from a force decrease to increase as compared to the rigid counterpart. \edit{We perform additional experiments with high stiffness flexible impactors that, despite constituting a less robust representation of a linear 2 DOF system, illustrate the behavior of the system in the limit of high hydroelastic factor.} A predictive theory is simultaneously developed that accounts for the added mass effect during the slamming stage, and is shown to quantitatively capture the measurements. \edit{As we demonstrate by considering the case of linear damping,} the simple theory can be readily extended to other nose geometries or structural elements and thus is anticipated to prove useful for the design and analysis of more complex engineered structures.

\section{Experimental methods}

\begin{figure}
    \centering
    \includegraphics[width = \textwidth]{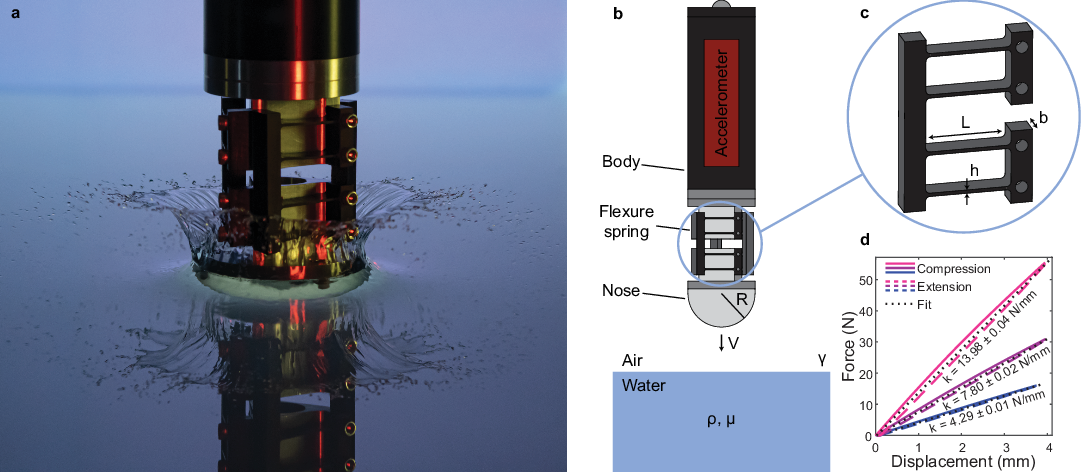}
    \caption{(a) Photograph of the flexible impactor ($k=$ 4.29 N/mm) entering the water at 2 m/s. An impact filmed with similar lighting and camera angle may be seen in Supplementary Movie 1. (b) Diagram of the flexible impactor with the main experimental parameters labeled. The rigid nose and body are connected by a set of three elastic flexure springs in a triangular configuration so the impactor system behaves like a simple harmonic oscillator. (c) Closeup of the flexure spring design. (d) Plot of force versus displacement for the elastic impactor with three different stiffness values. For each stiffness, we performed five trials whose standard deviation is smaller than the line width. Linear regression fitting to the combined compression and extension data is used to extract the reported linear stiffness values.}
    \label{fig:experimental_methods}
\end{figure}

\renewcommand{\arraystretch}{1.5} 
\edit{
\begin{table}
    \centering
    \caption{\edit{Relevant parameters and their range of values in our experimental study.}}
    \begin{tabular}{lccl}
        \textbf{\edit{Parameter}}              & \textbf{\edit{Symbol}}         & \textbf{\edit{Definition}}           & \textbf{\edit{Value}}                                                              \\
        \edit{Impact speed}                    & \edit{$V$}                     & \edit{--}                            & \edit{2 to 6 m/s}                                                              \\
        \edit{Nose radius}                     & \edit{$R$}                     & \edit{--}                            & \edit{22.23 or 29.64 mm}                                                           \\
        \edit{Flexible impactor stiffness}     & \edit{$k$}                     & \edit{--}                            & \edit{(low stiffness) 4.29, 7.80, 13.98 N/mm} \\
        \vspace{-0.68cm} \\
                                        &                           &                               & \edit{(high stiffness) 760, 21600 N/mm} \\ 
        \edit{Total flexible impactor mass}    & \edit{$M$}                     & \edit{--}                            & \edit{0.592 to 0.716 kg}                                                           \\
        \edit{Nose mass ratio}                 & \edit{$\alpha$}                & \edit{--}                            & \edit{(3D printed) 0.12} \\
         \vspace{-0.68cm}\\
                                        &                           &                               &         \edit{(aluminum) 0.24}\\ 
        \edit{Reynolds number}                 & \edit{$\textrm{Re}$}             & \edit{$\rho V R/\mu$}                  &      \edit{44,300 to 177,000}                         \\
        \edit{Weber number}                    & \edit{$\textrm{We}$}           & \edit{$\rho V^2 R/\gamma$}              &        \edit{1,220 to 14,600 }                    \\
        \edit{Froude number}                   & \edit{$\textrm{Fr}$}             &  \edit{$V^2/(g R)$}                       &     \edit{14 to 165}    \\
        \edit{Hydroelastic factor}             & \edit{$\textrm{R}_\textrm{F}$} & \edit{$\sqrt{\frac{k}{M \alpha (1-\alpha)}} \frac{R}{V}$} & \edit{0.68 to 185}                                              
    \end{tabular}
\label{tab:parameters}
\end{table}
}

\subsection{Experimental setup}

We perform experiments in which a slender impactor with a hemispherical nose (radius $R$ = 22.23 mm or 29.64 mm) enters a quiescent water bath with fluid density $\rho$, viscosity $\mu$, and interfacial tension $\gamma$.  Impacts are at normal incidence with impact speeds $V$ ranging from 2 to 6 m/s.  A schematic of the two degree-of-freedom impactor and bath is shown in figure \ref{fig:experimental_methods}(b) \edit{and the range of parameters in our experiments is reported in table \ref{tab:parameters}}. The typical Reynolds number $\rho V R/\mu$ is $O(10^5)$, the Weber number $\rho V^2 R/\gamma$ is $O(10^4)$ and the Froude number $V^2/(g R)$ is $O(10)$. Hence the fluid resistance of the presently studied impacts is dominated by fluid inertia, with additional resistances due to viscosity, surface tension, and hydrostatics as much weaker effects. The water bath is rectangular with length, width and depth of about 1 m, in order to approximate an infinite domain and avoid the influence of reflected surface waves during impact. The flexible impactor consists of a rigid body and nose coupled by a set of elastic flexure spring elements. The body contains an onboard accelerometer 
to measure the impact deceleration. A ferromagnetic ball is embedded in the body which allows the impactor to be dropped into free fall from an electromagnet at varying heights. Figure \ref{fig:experimental_methods}(c) shows the design of a typical flexure spring element which we laser-cut out of acetal plastic. The flexure features four thin beams which attach at the ends to a thick backbone and two mounting pads which are bolted to the impactor nose and body. The nose and body are coupled by three flexure spring elements in a rotationally symmetric pattern, as can be seen in figure \ref{fig:experimental_methods}(a). The flexures may be modeled as a set of guided cantilever beams \citep{judy1994flexures} and hence the overall axial stiffness $k$ of the flexible impactor can be estimated by $k = 3 E b h^3 / L^3$ where $E$ is the elastic modulus of the material and $b$, $h$ and $L$ are the beam dimensions. In practice, this equation overpredicts the impactor stiffness because it does not account for compliance of the flexure backbone or mounting pads.  Quasi-static compression testing of the flexures reveals a linear response with minimal hysteresis, as illustrated in figure \ref{fig:experimental_methods}(d), and we vary the flexure beam height $h$ in order to achieve the experimental stiffness values of $k=$ 4.29, 7.80 and 13.98 N/mm. \edit{Based on the results of preliminary experiments, these stiffness values were chosen in order to observe both a peak force decrease and increase as compared to the rigid case.} Due to the flexure configuration and beam geometry, the other bending and torsional modes of the impactor are much stiffer than the axial mode; in addition to behaving like springs, the flexures also serve the purpose of bearings which guide the relative motion of the nose and body. The key advantage of the flexure design is that, due to the monolithic structure, the axial elastic mode of the impactor experiences no static or sliding friction as it deflects. A photograph of the impactor during water entry in figure \ref{fig:experimental_methods}(a) \edit{(or Supplementary Movie 1)} shows the behavior of the flexures as the impactor achieves a submergence depth of approximately one radius and is enveloped in a crown splash. Despite the minimal deflection of the flexure beams, we find that the elasticity has a profound effect on the impact dynamics. \edit{Additional experiments are performed with high axial stiffness impactors by replacing the lower stiffness flexures with flexures that are significantly shortened ($k=$ 760 N/mm) or a solid plate of acetal plastic ($k=$ 21600 N/mm).  Although these configurations allow us to explore significantly higher axial stiffnesses while keeping all other parameters fixed, the axial mode is no longer the fundamental oscillation mode.}  Additional details regarding the impactor fabrication and characterization may be found in Appendix A. 

\subsection{Experimental procedure and processing}

The impactor is suspended from an electromagnet with the tip of the nose at a height $H \edit{=} V^2/(2g)$ above the air-water interface and allowed to rest for 5 minutes in order for any minimal swinging motion to decay. The impactor is then dropped into free fall while the onboard accelerometer (enDAQ S4) measures acceleration in three orthogonal axes (one axial, two radial) with 20 kHz sampling rate. Impacts are illuminated with diffuse white back light and filmed at 20,000 frames per second with a Phantom Veo camera equipped with a 50 mm Nikon lens at the height of the air-water interface or slightly above. The impact speed $V$ is measured from the high-speed entry videos by dividing the difference in the nose tip position at impact and 50 frames before impact by the appropriate time difference (2.5 ms). The impactor diameter is used as the length reference to convert from pixels to physical units. \edit{Because the field of view is centered on the impactor nose at the moment of impact, lens distortions have a negligible effect on the velocity measurements. Furthermore, we set the drop height $H$ using a precisely marked plumb line which results in a maximum impact velocity uncertainty of 2.5\% with respect to the nominal value.} In order to plot only the acceleration due to impact with the water surface and not the contribution from gravity, the mean accelerometer reading during free fall is subtracted from the acceleration data. When estimating the maximum impact acceleration, 
the slamming phase peaks in acceleration signals from each trial are first aligned using the cross-correlation method and $t=0$ is selected as the point at which the acceleration readings noticeably depart from free-fall behavior. Then a time window is defined which encompasses the maximum readings in the raw data from each trial. Estimates for the trial-averaged maximum acceleration 
and time of the peak  
use all of the data points in this time window. This method allows both sensor noise and the deviation between trials to be included in the uncertainty estimate on the peak. The error bars in 
subsequent figures incorporate this peak uncertainty as well as, when appropriate, other measurement uncertainty on values such as the impactor stiffness by use of the standard Taylor series approximation for multivariable error propagation from uncorrelated variables.

\section{Results}

\subsection{Peak deceleration of the flexible impactor}

\begin{figure}
    \centering
    \includegraphics[width = \textwidth]{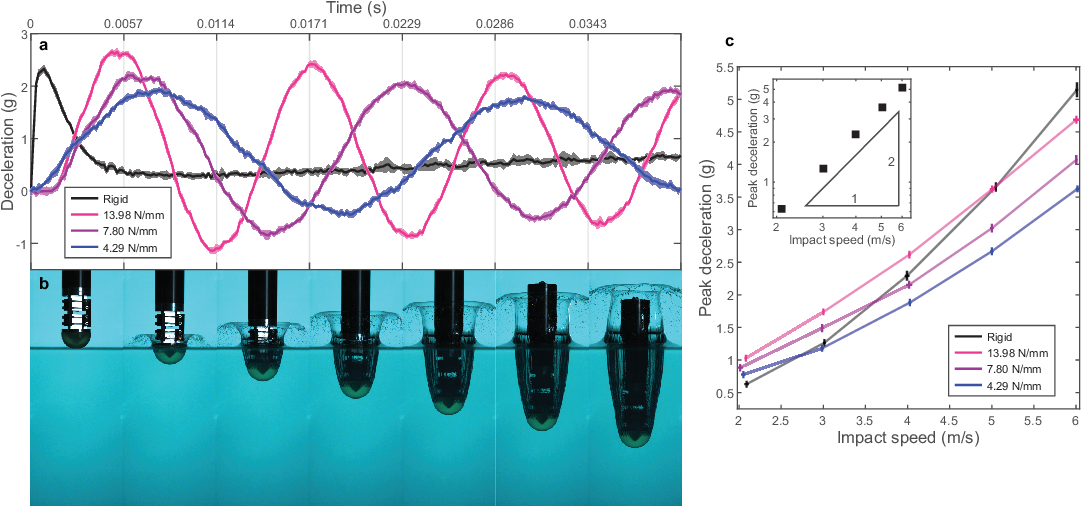}
    \caption{(a) Plots of impact deceleration versus time for the rigid \edit{($M=0.578$ kg)} and flexible (\edit{$M=0.592$ kg} and $k=$ 4.29, 7.80, 13.98 N/mm) cases with $V=$ 4 m/s and $R=$ 22.23 mm. The results are averaged over 5 trials and the shaded regions indicate the standard deviation between trials. \edit{The corresponding impactor velocities are presented in Appendix B}. (b) High-speed images of the flexible impactor ($k=$ 4.29 N/mm, $V=$ 4 m/s, $R=$ 22.23 mm) as it enters the water. The time of each photograph corresponds to the time axis label to the left of the image in (a).  A video version is available as Supplementary Movie \edit{3}. (c) Peak deceleration of the flexible impactor body depends on both stiffness and impact speed, so it can experience either a peak acceleration increase or decrease compared to the rigid case. The error bars show the standard deviation between 5 trials. \edit{The inset log-log plot of the rigid data shows that the peak acceleration increases like the square of the impact velocity.}}
    \label{fig:acceleration_results}
\end{figure}

Figure \ref{fig:acceleration_results}(a) shows the impactor body deceleration as a function of time after the moment of first impact for experiments with $V=$ 4 m/s and $R=$ 22.23 mm. As a baseline, we conduct experiments with an equivalent rigid impactor which has the same total mass \edit{(within 2.4\%)} as the flexible impactor. The rigid impactor is created by omitting the flexure assembly and rigidly fixing the nose to the impactor body.     The rigid impactor experiences a sharp peak force during early times in the slamming phase \edit{which occurs up to a submergence depth of approximately one radius}. However, despite the large forces during the slamming phase, the speed of the \edit{rigid} impactor \edit{decreases} only minimally -- \edit{on average 1.6\% by $t = R/V$ (one radius depth)} -- and it proceeds at a high rate into the bath, forming a crown splash and eventually a trailing air cavity as it pierces deeper into the water (Supplementary Movie \edit{2}). \edit{The impactor speed is plotted directly against time for these experimental cases in Appendix B.} The splash and cavity formation for the flexible impactor, as seen in figure 2(b) or Supplementary Movie \edit{3}, are similar to the rigid case \edit{during the slamming phase, though the cases with the 3D printed noses sometimes feature larger splashes and enhanced cavity size, likely due to the higher surface roughness and lower wettability compared to the machined aluminum noses \citep{watson2021wetting, speirs2019contactangle, aristoff2009hydrophobic, duez2007splash}. Increased hydrophobicity has also been shown to increase the force of impact on a sphere throughout the cavity-forming phase \citep{truscott2012unsteady}, though the effect is predominantly isolated to larger depths than focused on in the present work.} Although the interfacial physics of \edit{the cavity-forming phase} are certainly rich, the deceleration of the rigid impactor during the time after the slamming phase \edit{in our experiments} is relatively uninteresting, increasing only slowly as the steady state drag develops. The acceleration profiles of the flexible impactor body entering at 4 m/s with $k=$ 4.29 N/mm and 7.80 N/mm (figure \ref{fig:acceleration_results}(a)) follow the typical intuition for a ``cushioned'' impact; the impulse from the water entry ``shock'' is spread out over a longer time. Consequently, the maximum deceleration is decreased compared to the rigid case and the peak occurs at a later time, around one radius of submergence depth. The oscillations of the impactor persist for several cycles, dissipating slightly \edit{due to damping} as the impactor continues its descent. \edit{Like in the rigid case, the mean of the flexible impactor deceleration curve increases gradually at later times as the steady state drag develops.} However, surprisingly, the peak deceleration for the flexible impactor with $k=$ 13.98 N/mm is {\it higher} than the rigid case, indicating that, in general, the body acceleration during impact can either increase or decrease as a result of adding elasticity. Furthermore, we find that whether the impact acceleration increases or decreases for a given stiffness depends on the impact speed. The peak deceleration of the rigid and flexible impactors is plotted against impact speed in figure \ref{fig:acceleration_results}(c). For all experiments in this figure, the total impactor mass is held constant and, in the flexible impacts, the nose contains 12\% of the total mass. The peak deceleration of the rigid impactor increases like the square of the impact speed \edit{as shown in the inset in figure \ref{fig:acceleration_results}(c)}, confirming the anticipated inertially dominated regime. At the highest speed, all of the flexible impactors experience a peak deceleration reduction compared to the rigid case but the opposite is true at the lowest speed, where all of the flexible impactors experience a peak deceleration increase. This result portends an important subtlety in the design of elastic force reduction mechanisms for applications. Namely, the stiffness of the impactor must be carefully matched to the operating conditions or else the addition of elasticity can have significant deleterious effects.  In order to rationalize, interpret, and synthesize these observations, we develop a reduced mathematical model of the impact forces in what follows.

\subsection{Rigid added mass model}

\begin{figure}
    \centering
    \includegraphics[width = \textwidth]{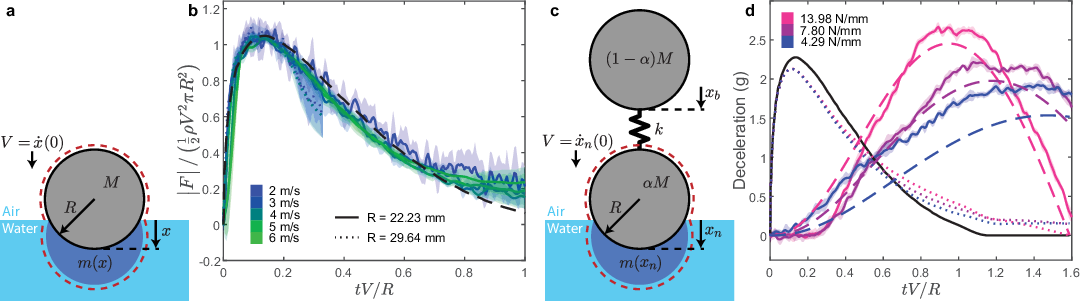}
    \caption{(a) Schematic of the rigid added mass model. The impinging body must accelerate an effective mass $m(x)$ of fluid as it enters the water; hence the impact force can be obtained from an expression of momentum conservation for the outlined system. (b) The force from rigid impact experiments at several speeds with different nose radii ($R=22.23$, $M=$ 0.578 kg or $R=29.64$, $M=$ 0.537 kg) collapse onto a single curve when using an inertial scaling. The Shiffman and Spencer \edit{$\textrm{C}_\textrm{F}$ curve} (dashed line) agrees excellently with the experiments. The shaded regions around the experimental curves indicate standard deviation of at least 3 trials \edit{with the lower speed experiments exhibiting greater variation between trials due to the lower signal-to-noise ratio}. The nose used for the $R=$ 29.64 mm rigid experiments is not a complete hemisphere so the data is truncated accordingly (importantly, the peak force is captured accurately). (c) We extend the classic added mass model to the case of a flexible impactor by introducing a trailing spring and mass. The outlined system for which we write conservation of momentum now includes the external contribution from the spring. (d) The deceleration of the flexible impactor body measured in experiments (solid colored lines, $V=$ 4 m/s, $R=$ 22.23 mm, $\alpha=$ 0.12, $M=$ 0.592 kg) is captured well by the flexible added mass model (dashed lines). Furthermore, the model predicts that the deceleration of the impactor center of mass (dotted lines) deviates only slightly from the rigid \edit{theoretical $\textrm{C}_\textrm{F}$ curve} (solid black line), suggesting that the elasticity does not have a strong influence on the hydrodynamics in this regime. The shaded region around the experimental curves indicates the standard deviation between 5 trials.}
    \label{fig:model}
\end{figure}

The hydrodynamic force during the slamming phase can be understood in the context of the added mass effect, originally applied by \cite{von1929impact} to the problem of water entry. Since the impact is inertially dominated, the impact force may be thought of as the rate at which the impactor must transfer momentum to a virtual quantity of fluid mass -- $m(x)$ in figure \ref{fig:model}(a) -- in order to accelerate that added fluid mass to the current impactor speed $\dot{x}(t)$. Conservation of momentum on the impactor and added mass system dictates that

\begin{equation}
    \frac{\textrm{d}}{\textrm{d} t} \left[(M+m)\dot{x} \right] = 0
\end{equation}
where $M$ is the mass of the impactor. Integrating with the initial conditions that $m(0) = 0$ and $\dot{x}(0)=V$ yields an expression for the acceleration during impact \citep{abrate2011hull}:

\begin{equation}
    \ddot{x} = - \frac{(M V)^2}{(M+m)^3} \frac{\textrm{d} m}{\textrm{d} x}.
    \label{eqn:rigid_accel}
\end{equation}

Thus the impact force profile can be calculated given only the added mass $m$ as a function of depth $x$. The added mass function for a sphere was derived by \cite{shiffman1945sphere1, shiffman1947lens} from potential flow around an axisymmetric lens. By neglecting deformation of the interface and higher order velocity terms, the dynamic boundary condition reduces to a statement of zero potential at $x=0$, and hence the impact problem is equivalent to uniform flow around the submerged portion of the body, mirrored about the undisturbed interface. Shiffman and Spencer later refined the model with additional theoretical and experimental corrections which account for the deformation of the interface and wetting of the sphere \citep{shiffman1945sphere2}. This corrected added mass function $m(x)$ is used in the present work \edit{and is reproduced along with its derivative $\textrm{d} m / \textrm{d} x$ in the supporting datasets. Although the corrected curve is only available from \cite{shiffman1945sphere2} up to a depth of $x=R$, we assume it smoothly tapers to zero at $x=1.15R$ and remains zero thereafter.  Whether the value of $\textrm{d} m / \textrm{d} x$ for $x>R$ is left constant, set abruptly to zero, or smoothly tapered to zero makes no significant quantitative difference in our predictions presented here.  These functions are the only externally derived components of our model, which is otherwise self-contained and described completely herein.} Shiffman and Spencer also define a dimensionless number

\begin{equation}
    \sigma = M / \left(\frac{4}{3} \pi R^3 \rho \right)
\end{equation}
which compares the impactor mass to the mass of an equivalent volume of fluid assuming a spherical impactor. When $\sigma$ is sufficiently large, as in the current experiments, $M \gg m$ (that is, the impactor mass is much larger than the peak added mass) and equation \ref{eqn:rigid_accel} can be reduced to

\begin{equation}
    F = -V^2 \frac{\textrm{d} m}{\textrm{d} x}
\end{equation}
where $F$ is the impact force on the body. As a consequence of large $\sigma$, the speed of the body does not change appreciably during the slamming phase and the impact force reaches a limiting curve (in practice, when $\sigma=3$, the maximum impact force is already 96\% of the infinite $\sigma$ case). Hence, we can define an impact drag coefficient $\textrm{C}_\textrm{F}$ based on $\textrm{d} m / \textrm{d} x$ as

\begin{equation}
    F(t) = \frac{1}{2} \textrm{C}_\textrm{F}(t) \rho V^2 \pi R^2.
    \label{eqn:F_vs_CF}
\end{equation}

 Assuming a heavy impactor with a given nose shape, $\textrm{C}_\textrm{F}$ is a function of time (or, interchangeably, depth) alone. The impact drag coefficient for the high $\sigma$ limit reported by \cite{shiffman1945sphere2} agrees excellently with rigid experiments performed with $V=$ 2 to 6 m/s and $R=$ 22.23 mm or 29.64 mm, as shown in figure \ref{fig:model}(b). The experimental curves collapse with the inertial force scale $\frac{1}{2} \rho V^2 \pi R^2$ and impact time scale $R/V$\edit{, indicating that the added mass during the slamming phase is independent of both the impactor speed and the size of the splash and air cavity, which vary throughout the experimental range of impact speeds}. \edit{Because the nose is not a complete hemisphere in the $R=29.64$ mm rigid experiments (it is a complete hemisphere in all other experiments), the force curves in figure \ref{fig:model}(b) are truncated at the point where the outer edge of the nose would first contact the undisturbed free surface. However, the peak force, which is the primary quantity of interest and used for comparison with the flexible experiments, is still captured accurately. Furthermore, the rigid results for both radii agree well with the numerous classic experimental studies for the force of impact on a rigid sphere in terms of both scaling and peak impact drag coefficient \citep{watanabe1934resistance, moghisi1981experimental, richardson1948impact, may1948drag, may1950virtualmass}.}

\subsection{Flexible added mass model}

We extend the added mass model to the case of an impacting simple harmonic oscillator by considering an external spring force on the nose-plus-added-mass system, as illustrated in figure \ref{fig:model}(c). The parameter $\alpha$ is introduced which equals the ratio of the nose mass to the total impactor mass, $M$. Hence conservation of momentum for the impactor nose plus added mass system is written as

\begin{equation}
    \frac{\textrm{d}}{\textrm{d}t} \left[ (\alpha M + m) \dot{x}_n \right] = k (x_b - x_n)
\end{equation}
where $x_b$ and $x_n$ are the positions of the impactor body and nose, respectively.    Integrating as before, the equation of motion for the nose is

\begin{multline}
    \ddot{x}_n = \frac{\textrm{d}}{\textrm{d}t} \left[ \frac{1}{\alpha M + m} \int_0^t k (x_b-x_n) \, \textrm{d} \tau \right] - \frac{\alpha^2 M^2 V^2}{(\alpha M + m)^3} \frac{\textrm{d}m}{\textrm{d}x_n}\\ - \frac{\alpha M V}{(\alpha M + m)^3} \frac{\textrm{d}m}{\textrm{d}x_n} \int_0^t k (x_b-x_n) \, \textrm{d} \tau.
    \label{eqn:nose_eom}
\end{multline}

The second term on the right hand side is equivalent to the hydrodynamic resistance from added mass as presented in equation \ref{eqn:rigid_accel}, while the other terms are new and account for the additional momentum exchange via the spring element.  Since the only force felt by the body comes from the spring, the equation of motion for the body is simply

\begin{equation}
    \ddot{x}_b = \frac{1}{(1-\alpha) M} k (x_n - x_b).
    \label{eqn:body_eom}
\end{equation}

Using the Shiffman and Spencer added mass function $m$, equations \ref{eqn:nose_eom} and \ref{eqn:body_eom} are numerically integrated with an RK4 scheme with trapezoidal rule for the integral terms and the deceleration of the body is compared with experimental results at $V =$ 4 m/s and $\alpha=0.12$ in figure \ref{fig:model}(d). The model agrees well with the experimental data and captures the transition in the peak deceleration for experiments with $k=$ 13.98 N/mm although it tends to underpredict the peak deceleration, most significantly for the $k=$ 4.29 N/mm experiments. In this case, the peak deceleration occurs after $tV/R=1$ in the region where Shiffman and Spencer's potential flow theory predicts that $\textrm{C}_\textrm{F}$ goes to zero. Experimentally, however, the contribution of form drag leads to non-zero impact force at late times as seen in figures \ref{fig:acceleration_results}(a) and \ref{fig:model}(b). When the peak deceleration occurs at later non-dimensional times (such as with low stiffness or high speed), the accuracy of the model can be improved by including the contribution of form drag -- directly from the experimental curves in \ref{fig:model}(b), for instance, as shown in Appendix \edit{C} -- in the added mass function $m$. Figure \ref{fig:model}(d) also shows the prediction for the deceleration of the flexible impactor center of mass, $\ddot{x}_c = \alpha \ddot{x}_n + (1-\alpha) \ddot{x}_b$, which is equal to the overall deceleration due to the hydrodynamic force. Despite its simplicity, this model captures the two-way coupling between the hydrodynamic force and impactor elasticity in the problem, \edit{with the added mass force term varying both explicitly as a function of nose depth and implicitly based on the nose and body velocities through the structural coupling}. However, the hydrodynamic force deviates only slightly from the rigid case, indicating that a further simplified one-way coupled model may be adequate. By formally assuming $\alpha M \gg m$ (that is, the nose mass is much larger than the peak added mass), equations \ref{eqn:nose_eom} and \ref{eqn:body_eom} simplify to

\begin{align}
    (1-\alpha) M \ddot{x}_b &= k (x_n - x_b) \label{eqn:body_one_way}, \\
    \alpha M \ddot{x}_n &= k (x_b - x_n) + F(t) \label{eqn:nose_one_way},
\end{align}
where $F(t)$ is the hydrodynamic force in the rigid case given by equation \ref{eqn:F_vs_CF}. In our experiments, the ratio $\alpha M / \max[m]$ takes values from 1.7 to 9.3. \edit{A solution to equations \ref{eqn:body_one_way} and \ref{eqn:nose_one_way} essentially comprises the structural response of the flexible impactor to the hydrodynamic forcing associated with the impact of a rigid sphere at constant velocity.}

\subsection{Peak force transition}

\begin{figure}
    \centering
    \includegraphics[width=\textwidth]{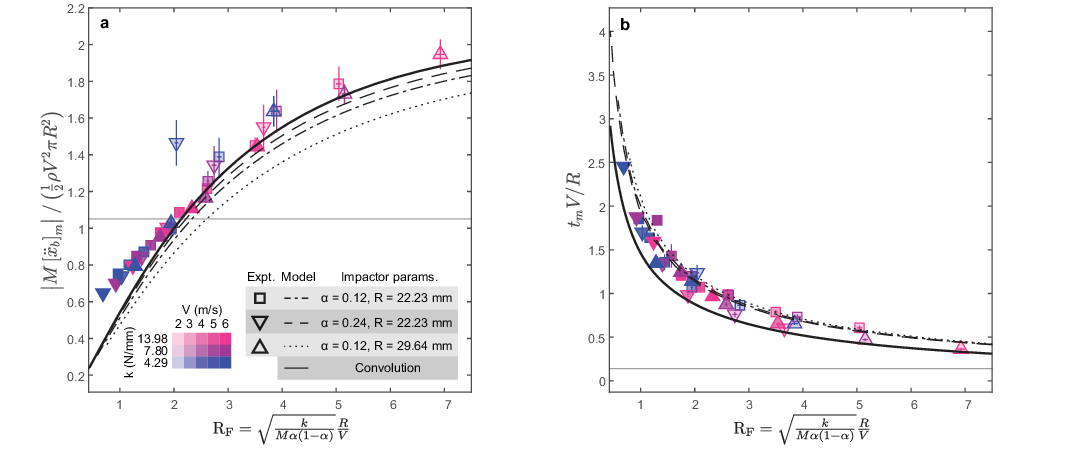}
    \caption{The scaled maximum impact force (a) and the time of the peak force (b) collapse along a single curve against the hydroelastic number $\textrm{R}_\textrm{F}$ for experiments in which the impact speed, stiffness, nose radius and mass ratio are varied. The error bars, which are sometimes smaller than the marker size, show the standard deviation between at least 3 trials.  The simplified prediction from the convolution integral in equation \ref{eqn:convolution} (solid black lines) agrees well with the experiments \edit{(markers)} and captures the critical hydroelastic factor near $\textrm{R}_\textrm{F}\approx 2$ at which the peak force in the flexible case equals the peak force in the equivalent rigid case (horizontal line). \edit{The marker shape indicates the impactor mass ratio and nose radius in a given experiment while the color and opacity indicate the stiffness and impact speed, respectively.} The two-way coupled added mass model from equations \ref{eqn:nose_eom} and \ref{eqn:body_eom} is also shown, which more accurately predicts the time of the peak force. \edit{The two-way model line style (dashed, dotted, or dash-dotted) indicates the mass ratio and nose radius corresponding to a particular predicted curve as shown in the legend.}}
    \label{fig:collapse}
\end{figure}

In order to understand the mechanism by which a high stiffness impactor can experience increased force compared to the equivalent rigid impactor, equations \ref{eqn:body_one_way} and \ref{eqn:nose_one_way} can be recast in modal coordinates, resulting in equations of motion for the rigid body mode (center of mass) and elastic mode as

\begin{align}
    \ddot{x}_c &= \frac{F(t)}{M}, \label{eqn:com_one_way} \\
    \ddot{\delta} + \frac{k}{\alpha (1-\alpha) M} \delta &= \frac{-F(t)}{\alpha M}, \label{eqn:mode_one_way}
\end{align}
where $\delta = x_b-x_n$. When $\delta=0$, the spring is at its natural length. By combining equations \ref{eqn:body_one_way} and \ref{eqn:mode_one_way}, the acceleration of the impactor body can be written as

\begin{equation}
    \ddot{x}_b = \frac{F(t)}{M} + \alpha \ddot{\delta}. \label{eqn:body_accel_contributions}
\end{equation}

Consequently, the acceleration of the impactor body (the quantity of interest, and directly measured in experiment), can be understood as the sum of contributions from the hydrodynamic force and the elastic mode. Furthermore, from equation \ref{eqn:mode_one_way}, it can be seen that $\ddot{\delta}>0$ in very early times (as $F(t)<0$), and thus the spring initially serves to isolate the body from the hydrodynamic forcing.  For all of the presently studied impacts, the hydrodynamic force has the same characteristic shape with a sharp increase to the peak followed by a slower decay (figure \ref{fig:model}(d) solid black and dotted lines). For impacts with low stiffness springs, the contribution from the elastic mode counteracts the peak in hydrodynamic force throughout the slamming phase and, as a result, the body experiences reduced peak deceleration. On the other hand, in the high stiffness case, the impactor body experiences the high frequency oscillations of the elastic mode (with $\ddot{\delta}<0$ earlier in the slamming phase) on top of the hydrodynamic forcing and thus the peak deceleration is increased. This interpretation suggests that the relationship between the hydrodynamic time scale and the impactor oscillation frequency plays a key role in determining whether the peak force will increase or decrease: if the elastic mode begins to oscillate before the hydrodynamic force decays, the impactor body will experience increased force. This ratio of time scales emerges directly when non-dimensionalizing the governing equations \ref{eqn:mode_one_way} and \ref{eqn:body_accel_contributions} using the length scale $R$ and the time scale $R/V$ in order to reach

\begin{align}
    \ddot{\tilde{x}}_b &= \frac{3}{8} \frac{\textrm{C}_\textrm{F}}{\sigma} + \alpha \ddot{\tilde{\delta}}, \\
    \ddot{\tilde{\delta}} + \textrm{R}_\textrm{F}^2 \tilde{\delta} &= - \frac{3}{8} \frac{\textrm{C}_\textrm{F}}{\alpha \sigma}, \label{eqn:rf_oscillator}
\end{align}
where the non-dimensional variables are marked with tildes. The new non-dimensional parameter $\textrm{R}_\textrm{F}$ is defined as

\begin{equation}
    \textrm{R}_\textrm{F} = \sqrt{\frac{k}{M \alpha (1-\alpha)}} \frac{R}{V}.
\end{equation}

This so-called hydroelastic factor is the ratio of the time scale of the hydrodynamic loading to the free fundamental oscillation period of the elastic impactor. Since equation \ref{eqn:rf_oscillator} has the same form as an undamped simple harmonic oscillator subjected to external forcing, the non-dimensional force on the impactor body may be predicted by computing the convolution of the hydrodynamic forcing and the elastic unit impulse response as

\begin{equation}
    \frac{M \ddot{x}_b}{\frac{1}{2} \rho V^2 \pi R^2} = \textrm{R}_\textrm{F} \int_0^{\tilde{t}} \textrm{C}_\textrm{F} (\tau) \sin (\textrm{R}_\textrm{F} (\tilde{t} - \tau)) \, \textrm{d} \tau.\label{eqn:convolution}
\end{equation}

Figures \ref{fig:collapse}(a) and (b) show the non-dimensional maximum impact force $M \max[\ddot{x}_b]=M [\ddot{x}_b]_m$ and time of peak force $t_\textrm{m}$ as a function of $\textrm{R}_\textrm{F}$ for flexible impact experiments with $V=$ 2 to 6 m/s, $k=$ 4.29 to 13.98 N/mm, $\alpha=$ 0.12 or 0.24, and $R=$ 22.23 mm or 29.64 mm. The experiments collapse to a single curve which agrees excellently with the model in equation \ref{eqn:convolution}. At some critical value near $\textrm{R}_\textrm{F} \approx 2$, the peak force on the flexible impactor exceeds the non-dimensional peak force on the equivalent spherical rigid impactor which is approximately 1.05. The results of the two-way coupled model in equations \ref{eqn:nose_eom} and \ref{eqn:body_eom} are plotted as well; as previously discussed, the two-way coupled model tends to underpredict the maximum impact force but captures the time of peak force more accurately than the convolution integral. Thus, we have shown that the force on the body depends only on the impact drag coefficient $\textrm{C}_\textrm{F}$ for the equivalent rigid case, which is a consequence of the nose geometry, and the hydroelastic factor $\textrm{R}_\textrm{F}$, which depends on the design of the elastic structure and impact velocity. 

Two notable comments remain. The first regards the data point which furthest deviates from the theory in figure \ref{fig:collapse}(a), corresponding to experiments with $V=$ 2 m/s, $k=$ 4.29 N/mm and $\alpha=0.24$. The elastic mode of the flexible impactor is not constrained during free fall but, in all other cases, the oscillations are so slight that they have no noticeable effect on the peak acceleration. However, this data point represents the ``worst'' case scenario with the heaviest nose, weakest spring and least time during free fall for the oscillations to dissipate. As such, we observe that the phase of the free-fall oscillation at impact significantly influences the force. In particular, $\delta(0)<0$ for these experiments -- the impactor is ``pre-stretched'' -- which increases the impact force compared to the $\delta(0)=0$ case. This effect can be captured by changing the initial conditions of the two-way coupled model. Conversely, according to the model, a ``pre-compressed'' ($\delta(0)>0$) impactor would experience reduced force at these operating parameters. \edit{This point is explored in more detail in Appendix D.} The second comment is a reminder that only the slamming phase of impact is considered in the present study; in some rare cases (particularly at low $k$ or low $V$), the maximum force during slamming is exceeded at much later times when the trailing air cavity pinches off. Consequently, additional considerations must be made in these cases if the goal is to predict the maximum force during the entire impact.

\subsection{High stiffness limit}

\begin{figure}
    \centering
    \includegraphics[width=\textwidth]{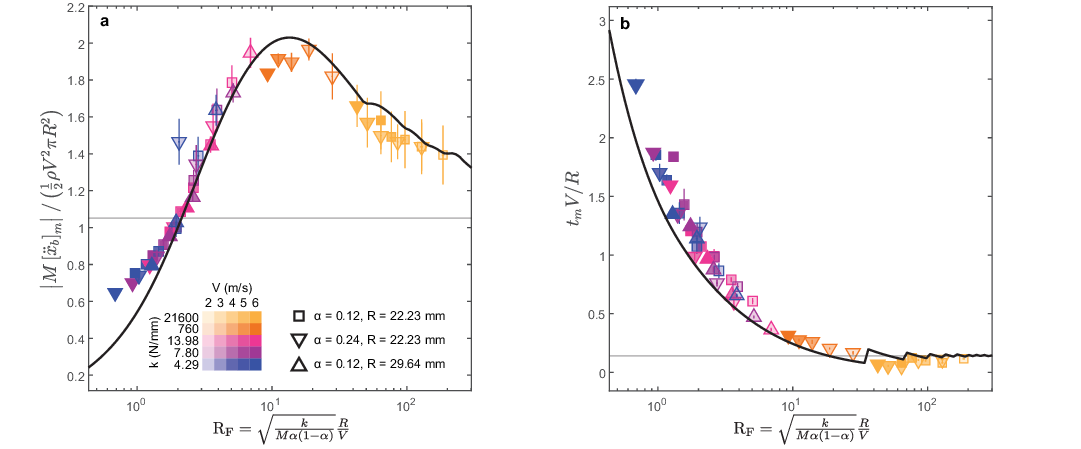}
    \caption{Additional experiments with two high stiffness impactors demonstrate that the scaled maximum impact force (a) and the time of the peak force (b) approach the equivalent rigid case (horizontal line) as $\textrm{R}_\textrm{F}$ becomes large. While the time of the peak effectively returns to the rigid case past $\textrm{R}_\textrm{F}\approx \edit{20}$, the non-dimensional peak force still exceeds the rigid case by more than 30\% at $\textrm{R}_\textrm{F}\approx 200$. The simplified prediction from the convolution integral (solid black line) accurately captures the impactor behavior at large $\textrm{R}_\textrm{F}$. The error bars, which are sometimes smaller than the marker size, show the standard deviation between at least 3 trials.}
    \label{fig:high_rf}
\end{figure}

Intuitively, as the stiffness of the impactor increases to sufficiently large values \edit{(corresponding to high $\textrm{R}_\textrm{F}$)}, its behavior should eventually return to the ``rigid'' case. In order to observe this behavior, we conducted experiments with two additional flexure spring designs with high stiffness values of $k=$ 760 N/mm and 21600 N/mm. Since these values exceed the capabilities of our tensile testing machine, the stiffness values were instead extracted from the impactor natural frequencies which were measured by suspending the impactor from a low stiffness bungee and exciting the axial mode with an impact hammer. \edit{As shown in figure \ref{fig:high_rf},} the high stiffness experiments \edit{performed with $V=2$ to 6 m/s, $R=$ 22.23 mm and $\alpha = 0.12$ or 0.24} also collapse nicely onto the theoretical curve predicted by equation \ref{eqn:convolution} \edit{despite the fact that the axial mode no longer represents the fundamental mode, with bending modes predicted to occur at lower frequencies (Appendix A)}. \edit{By substantially increasing the impactor's axial stiffness, it is possible to extend our experimental $\textrm{R}_\textrm{F}$ values by more than an order of magnitude while keeping other experimental parameters the same.} For impacts near $\textrm{R}_\textrm{F}\approx 10$, the non-dimensional peak force achieves values that are approximately double those in the equivalent rigid case before steadily decreasing as $\textrm{R}_\textrm{F}$ is further increased. However, despite spanning nearly four orders of magnitude in stiffness in our flexible experiments, the peak of the equivalent rigid case was not recovered: even at $\textrm{R}_\textrm{F}\approx 200$, the measured (and predicted) peak force exceeds the rigid case by more than 30\% \edit{as seen in figure \ref{fig:high_rf}(a)}. In fact, according to the model, it is not until $\textrm{R}_\textrm{F}$ achieves a value near 4,500 that the peak force returns to within 5\% of the rigid case. Past $\textrm{R}_\textrm{F}\approx \edit{20}$, the time of the peak force is essentially the same as in the equivalent rigid case \edit{as demonstrated in figure \ref{fig:high_rf}(b)}. Note that the odd bumpiness in the theoretical curves at high $\textrm{R}_\textrm{F}$ is physical in origin -- once $\textrm{R}_\textrm{F}$ is high enough for several impactor oscillations to occur before the hydrodynamic force reaches its peak, the time and magnitude of the peak force are highly sensitive to the relative phase of the impactor oscillations and hydrodynamic force. 

\subsection{Effect of damping}

\begin{figure}
    \centering
    \includegraphics[scale = 0.8]{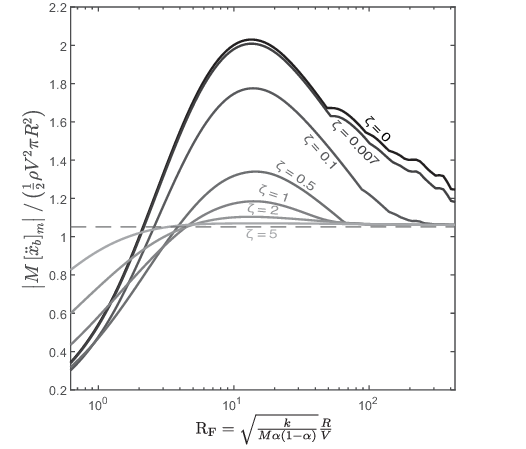}
    \caption{The theoretical prediction for the non-dimensional maximum impact force on a linearly damped impactor is plotted as a function of hydroelastic number $\textrm{R}_\textrm{F}$ and compared to the undamped case ($\zeta=0$) and the equivalent rigid impactor (dashed line). The $\zeta=0.007$ curve corresponds to the measured damping ratio of the flexible impactor design and differs minimally from the undamped case. An underdamped impactor could still experience substantial peak force reduction at low $\textrm{R}_\textrm{F}$ without incurring such a large peak force increase at large $\textrm{R}_\textrm{F}$ as compared to the undamped case.}
    \label{fig:damping}
\end{figure}

A final question of high practical relevance that can now be addressed with our validated model is the influence of damping on the behaviors elucidated herein. The model culminating in equation \ref{eqn:convolution} can be readily extended to an impactor which is a linearly damped harmonic oscillator. We replace the undamped unit impulse response function in the convolution with the appropriate damped unit impulse response function $g(t)$ and let

\begin{equation}
    I = \int_0^{\tilde{t}} \textrm{C}_\textrm{F}(\tau) g(\tilde{t} - \tau) \, \textrm{d}\tau. \label{eqn:damped_conv_integral}
\end{equation}

The unit impulse response functions $g(t)$ for the damped impactor used in equation \ref{eqn:damped_conv_integral} are solutions to

\begin{equation}
    \ddot{\tilde{\delta}} + 2 \zeta \textrm{R}_\textrm{F} \dot{\tilde{\delta}} + \textrm{R}_\textrm{F}^2 \tilde{\delta} = 0 \qquad \tilde{\delta}(0)=0 \qquad \dot{\tilde{\delta}}(0)=1
\end{equation}
which are given by
\begin{align}
    g(\tilde{t}) &= \frac{1}{\textrm{R}_\textrm{F}}\sin\left(\textrm{R}_\textrm{F} \tilde{t} \right) &\qquad \textrm{undamped ($\zeta=0$)}\\
    g(\tilde{t}) &= \frac{1}{\textrm{R}_\textrm{F}\sqrt{1-\zeta^2}} \exp \left(-\zeta \textrm{R}_\textrm{F} \tilde{t}\right) \sin\left(\textrm{R}_\textrm{F} \sqrt{1-\zeta^2} \tilde{t} \right) &\qquad \textrm{underdamped ($\zeta<1$)}\\
    g(\tilde{t}) &= \tilde{t} \exp \left(-\textrm{R}_\textrm{F} \tilde{t} \right) &\qquad \textrm{crit. damped ($\zeta=1$)}\\
    g(\tilde{t}) &= \frac{1}{2 \textrm{R}_\textrm{F}\sqrt{\zeta^2-1}}\left[ e^{\textrm{R}_\textrm{F} \tilde{t}\left(\sqrt{\zeta^2-1}-\zeta \right) } - e^{-\textrm{R}_\textrm{F} \tilde{t}\left(\sqrt{\zeta^2-1}+\zeta \right) } \right] &\qquad \textrm{overdamped ($\zeta>1$)}
\end{align}

Then, the dimensionless impact force on the body is given by

\begin{equation}
    \frac{M \ddot{x}_b}{\frac{1}{2} \rho V^2 \pi R^2} = \textrm{R}_\textrm{F}^2 I + 2 \zeta \textrm{R}_\textrm{F} \dot{I}. \label{eqn:damping}
\end{equation}

Here, $\zeta$ is the impactor damping ratio defined as

\begin{equation}
    \zeta = \frac{c}{2 \sqrt{M k \alpha (1-\alpha)}} 
\end{equation}
where $c$ is the damping coefficent. The theoretical maximum impact force versus $\textrm{R}_\textrm{F}$ is plotted for several values of $\zeta$ in figure \ref{fig:damping}. For the impactors with stiffness values $k=$ 4.29, 7.80 and 13.98 N/mm, the damping ratio $\zeta$ is 0.007 $\pm$ 0.001 as measured from ring down tests. Equation \ref{eqn:damping} predicts that damping tends to lower the peak force at high $\textrm{R}_\textrm{F}$, but increase the peak force at low $\textrm{R}_\textrm{F}$. This general trend is consistent with the physics of passive vibration isolation.  As $\zeta$ increases, the critical $\textrm{R}_\textrm{F}$ where the peak force is equal to the equivalent rigid case shifts in a non-monotonic way, first increasing and then starting to decrease near critical damping. At high damping, the curve returns to the rigid case. For an impactor \edit{with a fixed stiffness} that must perform over a wide range of operating conditions, an underdamped system could be designed to maintain significant peak force reduction at high speeds (low $\textrm{R}_\textrm{F}$) without incurring such a large penalty at low speeds (high $\textrm{R}_\textrm{F}$) as the undamped system.

\section{Discussion}

In the present work, we have provided experimental and theoretical treatment of a simplified hydroelastic problem involving the water entry of a 2DOF (one axial elastic mode) impactor with a hemispherical nose. The impactor nose and body are coupled with a set of compliant flexure springs in order to achieve a system that closely approximates a simple harmonic oscillator. Using an onboard accelerometer, we directly measure the deceleration of the body during water entry over a wide range of impact speeds, spring stiffnesses, nose radii and nose to body mass ratios. We accompany these experiments with a predictive theory based on the added mass effect and show that, in a certain regime with high practical relevance, the peak slamming force only depends on two dimensionless quantities. These quantities are the impact drag coefficient function $\textrm{C}_\textrm{F}$, which is prescribed by the nose geometry, and the hydroelastic factor $\textrm{R}_\textrm{F}$, which relates the hydrodynamic and impactor elastic mode time scales. At low $\textrm{R}_\textrm{F}$, which corresponds to low stiffness or high impact speed, the flexible impactor body experiences reduced force as compared to an equivalent rigid impactor. However, flexibility can also increase the force once $\textrm{R}_\textrm{F}$ exceeds a critical value. We use our validated model to make a prediction for the damped case and show that, if an impactor must operate over a wide range of $\textrm{R}_\textrm{F}$ values, an underdamped design could recover significant force reduction at low $\textrm{R}_\textrm{F}$ without such a penalty at high $\textrm{R}_\textrm{F}$. As reviewed in the introduction, the present work is by no means the first to consider the role of impactor elasticity on the structural loading during water entry.  Throughout prior studies, some have demonstrated systematic force reduction \citep{wu2020water} or force increase \citep{may1970review,shi2019numerical}, while others have suggested both possibilities depending on the timescales of the problem \citep{kim1996coupled,carcaterra2004hydrodynamic}, as corroborated herein.  However, despite the variety of reduced-order models available, very few experimental realizations of equivalently simplified structures have been completed to date.  Our integrated experimental and theoretical study of what is arguably the simplest possible hydroelastic problem has illuminated the essential fundamental physics while showcasing its richness, and may potentially serve as a foundation for more advanced studies in the field.

In particular, the present work has focused on an undamped simple harmonic oscillator with a hemispherical nose to minimize the total number of parameters in the problem.  However, both the one- and two-way coupled theoretical formulations presented can be immediately applied to a much broader range of problems.  For instance, the one-way coupled model (equation \ref{eqn:convolution}) can be easily extended to other nose geometries since it requires as input only the knowledge of the rigid impactor dynamics.  Given an arbitrary axisymmetric nose geometry, one may perform experiments with a rigid impactor to obtain $\textrm{C}_\textrm{F}$ and substitute the result into equation \ref{eqn:convolution} to predict the peak force for an elastic design.  While we have used theoretical results available for rigid spherical impactors \citep{shiffman1945sphere2}, equally good (or better) predictions can be made by directly using a fit to the collapsed data presented in figure \ref{fig:model}(b).  Furthermore, by suitably adapting the more general ODE formulation presented in equations \ref{eqn:nose_eom} and \ref{eqn:body_eom}, one can make predictions for non-linear structural elements or explore cases outside of the ``heavy'' nose limit where explicit two-way coupling plays a more prominent role.  Continuous structures may also be analyzed in the present framework using standard lumped-mass modeling approaches, and in many instances very few modes need to be resolved for faithful representations of the dynamics of otherwise complex structures \citep{piro2013hydroelastic}.  While significant progress has been made in high-fidelity coupled simulations of hydroelastic problems over the past several years, the reduced-order models presented herein are (in contrast) extraordinarily efficient to compute and may be particularly valuable in early design phases and in interpreting experimental and computational results.   

\vspace{3mm}
{\small 
\noindent {\bf Supplementary material.} Supplementary videos are made available to the interested reader. \\
\noindent {\bf Data accessibility.} The authors declare that all data supporting the findings of this study are available within the paper and supplementary datasets at \url{https://github.com/harrislab-brown/WaterEntrySHO}. Any additional data is available from the corresponding author on reasonable request.\\
\noindent {\bf Funding.} J.T.A. and D.M.H. acknowledge funding from the Office of Naval Research (ONR N00014-21-1-2816).  J.B. and N.B.S. acknowledge funding from the Naval Undersea Warfare Center In-House Laboratory Independent Research program, monitored by Dr. Elizabeth Magliula. \\
\noindent {\bf Acknowledgements.} J.T.A. and D.M.H. acknowledge the support of the Brown Design Workshop and JEPIS staff for use of their tensile testing equipment and advice regarding impactor fabrication. \edit{We also thank Pradeep Guduru and members of the Experimental Solid Mechanics Laboratory for advice regarding stiffness testing.} \\
{\bf Declaration of interests.} The authors report no conflict of interest.}

\section*{Appendix A. Impactor details}

\subsection*{Impactor fabrication}

The slender flexible axisymmetric impactor is designed with an overall diameter of 44.45 mm and length of approximately 220 mm. All parts are machined from ASTM 6061-T6 aluminum except for the flexure spring elements which are laser-cut from acetal plastic and the noses for experiments with $\alpha=0.12$, which are 3D printed out of photopolymer resin (Formlabs Clear Resin V4) on a Formlabs Form V2 machine. All aluminum components are anodized black according to MIL 8625 except for the parts which hold the flexure spring elements which are left uncoated to aid visualization. The impactor body contains a cavity into which the accelerometer is rigidly mounted using a threaded locking ring. An end cap with an O-ring ensures the body is water tight and contains an embedded steel sphere for dropping from the electromagnet. The flexure spring assembly features three flexure spring elements as shown in figure \ref{fig:experimental_methods}(c) which are bolted to triangular aluminum components. For the experiments with $k=$ 4.29, 7.80 and 13.98 N/mm, the flexure beam length $L$ is 18 mm, the material thickness $b$ is 6.78 mm and the beam height $h$ is varied from approximately 0.5 - 1.5 mm to achieve the different stiffness values. Fillets with 1 mm radius are applied at the beam connection points in order to reduce the stress concentration and enable the flexures to survive the large shock as the impactor collides with the bottom of the tank. For the high stiffness experiments with $k=$ 760 N/mm, the flexure beams are significantly shortened with $h/L \approx 1/3$. For the $k=$ 21600 N/mm experiments, the flexure elements are simply a solid plate of acetal plastic with the mounting holes cut out. The flexure spring assembly threads into the impactor body at one end and bolts to the nose at the other end. Hemispherical nose \edit{pieces} weighing \edit{0.029} kg are 3D printed with $R=$ 22.23 and 29.64 mm for the $\alpha=0.12$ experiments. An aluminum nose \edit{piece} weighing \edit{0.126} kg with $R=$ 22.23 mm is fabricated for the $\sigma=0.24$ experiments. \edit{The nose pieces are each attached to an aluminum carrier weighing 0.033 kg which features mounting holes for the flexure elements.} The fully assembled impactor body weighs \edit{0.514} kg and the flexure elements typically weigh 0.020 kg \edit{for the low stiffness experiments and 0.040 kg for the high stiffness experiments}, which we assume is split evenly between \edit{the body and nose masses}. \edit{The slight mass variations due to the flexure elements are taken into account when calculating the $\textrm{R}_\textrm{F}$ value and non-dimensional maximum impact force for a given experiment}. For the equivalent rigid experiments, additional aluminum noses are fabricated which thread directly into the impactor body in place of the flexure spring assembly. When possible, the lengths of the rigid noses are chosen so that the total impactor mass is the same as in the flexible experiments.

\subsection*{Characterization of impactor flexibility}

The flexible impactors with $k=$ 4.29, 7.80 and 13.98 N/mm are tested in an Instron 5924 tensile testing machine \edit{with a 500 N load cell} at a displacement rate of 1 mm/s to a total displacement of 4 mm. Force is measured during both the compression and subsequent extension in order to observe the hysteresis of the structure. The stiffness values $k$ are determined by linear regression fitting of the force data. \edit{For the high stiffness cases with $k=$ 760 and 21600 N/mm, an accurate displacement curve cannot be directly obtained from the Instron machine because the impactor stiffness is comparable to the machine frame stiffness;} instead the stiffness is estimated through experimental modal testing. The impactor is hung from a weak bungee such that the vertical translation mode has a low frequency (0.6 Hz) and the axial flexible mode of the impactor is excited by applying a longitudinal impulse with a rubber-tipped hammer. The acceleration of the impactor body is recorded and the natural frequency -- from which the axial mode stiffness is calculated -- is extracted by fitting a damped sinusoid to the data. This setup is also used with the low stiffness impactors ($k=$ 4.29, 7.80 and 13.98 N/mm) and the decay rate of the fit sinusoid  is used to estimate the damping ratio: $\zeta = 0.007 \pm 0.001$. The natural frequencies of the low stiffness impactors obtained from experimental modal testing suggest slightly higher stiffness values than obtained via quasi-static testing, possibly due to rate dependent material behavior, which would help to explain the underprediction of the two-way coupled model in figure \ref{fig:collapse}(a). However, we choose to report the values from quasi-static testing since this is a more direct and independent measurement of the stiffness. Computational modal analysis simulations are completed in Autodesk Fusion 360 and used to inform the design of the low stiffness impactors so that only the axial mode is excited during impact and they behave like simple harmonic oscillators. For the flexible impactors with $k=$ 4.29, 7.80 and 13.98 N/mm, the simulations predict that the first harmonics (bending of the flexure springs) always have at least 142\% higher natural frequency than the axial fundamental mode. The simplified flexure beam theory model \citep{judy1994flexures} supports this finding; the translational stiffness of the flexure element corresponding to tension/compression of the beams is larger than the axial mode stiffness by a factor which scales like $L^2/h^2$. Similarly, the translational stiffness of the flexure element corresponding to the other bending mode of its beams is larger by a factor which scales like $b^2/h^2$. For the high stiffness impactors ($k=$ 760 and 21600 N/mm), the design is less rigorous as the fundamental mode involves bending of the flexures. Nevertheless, the acceleration data during impact confirms that the axial mode is primarily excited due to the axial nature of the loading: the maximum off-axis vibrations during impact (measured in radial directions) always remain less than 21\% of the maximum axial acceleration, although are typically no more than 5\%.  

\section*{\edit{Appendix B. Impactor speed}}

\begin{figure}
    \centering
    \includegraphics[scale=0.8]{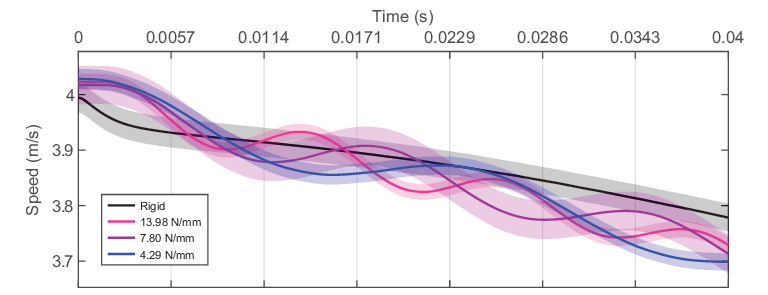}
\caption{\edit{Plots of impactor speed versus time for the rigid ($M=0.578$ kg) and flexible ($M=0.592$ kg; $k=$ 4.29, 7.80, 13.98 N/mm; $\alpha=0.12$) cases with $V=$ 4 m/s and $R=$ 22.23 mm obtained by integrating the deceleration data reported in figure \ref{fig:acceleration_results}(a). The results are averaged over 5 experimental trials and the shaded regions indicate the standard deviation between trials. Despite the large forces during the slamming phase, the changes to the impactor speed are relatively small due to its short duration.}}
    \label{fig:speed_vs_time}
\end{figure}

\edit{We report the impactor speed as a function of time in figure \ref{fig:speed_vs_time} for rigid and flexible experiments with $V=$ 4 m/s and $R=$ 22.23 mm. For the flexible cases, $k=$ 4.29, 7.80 and 13.98 N/mm, and $\alpha=0.12$. The impactor speed is obtained by integrating the acceleration data in figure \ref{fig:acceleration_results}(a) with a trapezoidal rule and the initial speeds from the camera measurements. Since the acceleration data is quite repeatable, the variation between trials in figure \ref{fig:speed_vs_time} is mainly attributed to the variation in measured impactor speed at the moment of impact.}  

\section*{Appendix \edit{C}. Influence of form drag}

\begin{figure}
    \centering
    \includegraphics[width=\textwidth]{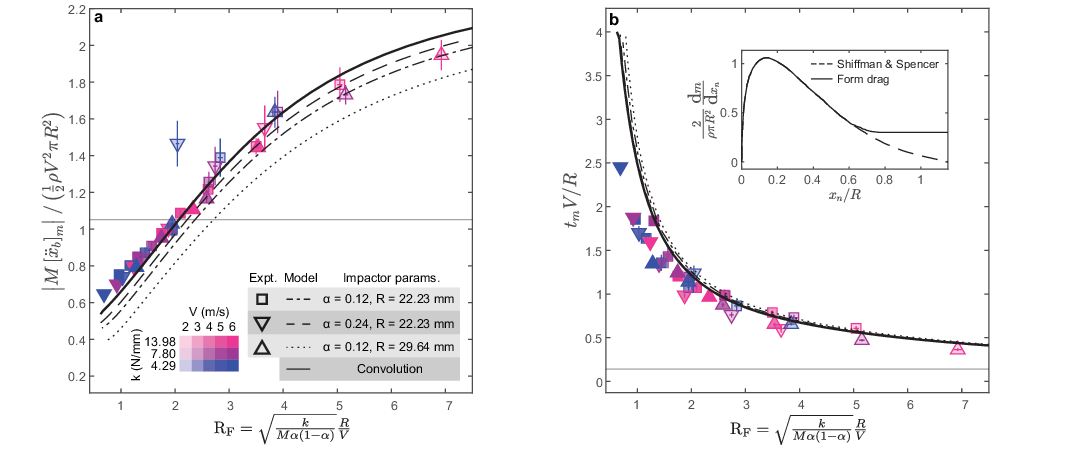}
    \caption{The scaled maximum impact force (a) and the time of the peak force (b) are plotted against the hydroelastic factor $\textrm{R}_\textrm{F}$ and compared with our model which is updated to include form drag. The error bars, which are sometimes smaller than the marker size, show the standard deviation between at least 3 \edit{experimental} trials. \edit{The marker shape indicates the impactor mass ratio and nose radius in a given experiment while
    the color and opacity indicate the stiffness and impact speed, respectively.} The simplified prediction from the convolution integral in equation \ref{eqn:convolution} is shown in the solid black lines and the two-way coupled added mass model from equations \ref{eqn:nose_eom} and \ref{eqn:body_eom} is also shown. \edit{The two-way model line style (dashed, dotted, or dash-dotted) indicates the mass ratio and nose radius corresponding to a particular predicted curve as shown in the legend.} The agreement with the theory improves compared to figure \ref{fig:collapse} when we update the added mass function $\textrm{d}m/\textrm{d}x$ to include the contribution of form drag as shown in the inset plot in (b).}
    \label{fig:form_drag}
\end{figure}

We modify the Shiffman and Spencer added mass function $\textrm{d}m/\textrm{d}x$ as shown in the inset plot in figure \ref{fig:form_drag}(b) in order to match the experimental impact force profile for the rigid impactor (figure \ref{fig:model}(b)) at later non-dimensional times and hence account for the contribution of form drag. This change improves the agreement between the predicted and experimentally measured peak impact acceleration as shown in figure \ref{fig:form_drag}(a), particularly at low $\textrm{R}_\textrm{F}$. The change in the prediction for the time of the peak acceleration is less pronounced, as shown in figure \ref{fig:form_drag}(b).

\section*{\edit{Appendix D. Influence of impactor pre-load}}

\begin{figure}
    \centering
    \includegraphics[scale=0.8]{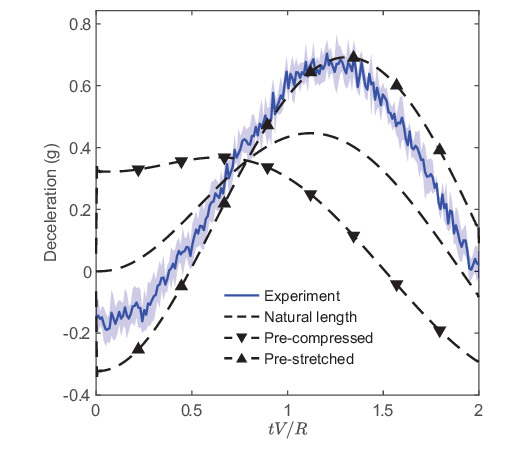}
    \caption{\edit{The experimental acceleration curve (solid blue line) with $\alpha=0.24$, $V=2$ m/s and $k=4.29$ N/mm is compared to the theoretical prediction of the two-way flexible added mass model (dashed lines). The shaded region represents the standard deviation of 3 experimental trials. The natural length curve corresponds to the impact model with no pre-load while the pre-stretched or pre-compressed curves are generated by modifying the initial conditions of the model with $\delta(0)=\pm 0.39$ mm and $\dot{\delta}(0)=0$. This initial displacement corresponds to the estimated free-fall oscillation amplitude based on the weight of the hanging nose and spring stiffness. The experimental curve closely matches the pre-stretched case suggesting that $\delta(0)<0$ for these experiments. The peak deceleration is notably increased for this case compared to the theoretical curve with no pre-load.}}
    \label{fig:preload}
\end{figure}

\edit{Since the elastic mode of the impactor is not constrained during free fall, the impactor can experience a pre-loaded impact in cases where the free-fall oscillations persist at the moment of impact such that $\delta(0) \neq 0$. Although we cannot reliably measure the nose displacement in our current experimental setup, we expect that at the moment of impact, the nose may be stretched or compressed by a maximum distance associated with its initial gravitational extension}
\begin{equation}
    \edit{\delta_\mathrm{grav} = \frac{\alpha M g}{k}.}
\end{equation}

\edit{Additional compression that might arise from aerodynamic effects during free fall are negligible for our experimental parameters. For the experimental case in which we observe the most significant effects due to the oscillation phase at impact ($V=2$ m/s, $\alpha=0.24$, $k=4.29$ N/mm), the displacement due to the nose weight is $\delta_\mathrm{grav}=0.39$ mm. Hence, assuming no damping during the free fall, in the most ``extreme'' cases the nose would have a positive or negative displacement of 0.39 mm at the moment of impact, with zero nose velocity relative to the body. By appropriately modifying the initial conditions based on $\delta(0) = \pm 0.39$ mm and $\dot{\delta}(0)=0$, we use the flexible added mass model in equations \ref{eqn:nose_eom} and \ref{eqn:body_eom} to predict the impact acceleration and compare to the experimental data for the case of $\alpha=0.24$, $V=2$ m/s and $k=4.29$ N/mm. The resulting acceleration curves in figure \ref{fig:preload} successfully bound the experimental data, and demonstrate that the predicted peak deceleration depends on the pre-load. For these parameters, the pre-stretched case experiences increased deceleration during impact while the pre-compressed case experiences reduced deceleration. The experimental curve more closely matches the pre-stretched case suggesting that $\delta(0)<0$ for these experiments.}

\bibliographystyle{jfm}
\bibliography{references}
\end{document}